\documentclass[aps,prb,twocolumn,preprintnumbers,showkeys,letterpaper,groupaddress]{revtex4}
\usepackage{amsfonts}
\usepackage{amsmath}
\usepackage{amssymb}
\usepackage{algorithm}
\usepackage{algorithmic}
\usepackage{bm}
\usepackage{dsfont}
\usepackage{color}
\usepackage{newlfont}
\usepackage{prettyref}
\usepackage[hyperfootnotes,hyperindex,hyperfigures]{hyperref}
\hypersetup{colorlinks=true, linkcolor=blue, citecolor=grass} 
\usepackage{mathrsfs}
\usepackage{program}
\usepackage{graphicx}

\providecommand{\abs}[1]{\lvert#1\rvert}

\providecommand{\norm}[1]{\lVert#1\rVert}

\providecommand{\Liou}{\mathds{L}}

\providecommand{\mtrx}[1]{$\bm #1 $}

\DeclareMathOperator*{\argmin}{argmin}
\definecolor{lightyellow}{rgb}{1.0,1.0,0.5}
\definecolor{peach}{rgb}{1.0,.8,.7}
\definecolor{khaki}{rgb}{0.9,0.9,0.7}
\definecolor{navy}{rgb}{0.0,0.1,0.7}
\definecolor{maroon}{rgb}{0.7,0.1,0.2}
\definecolor{teal}{rgb}{0.0,0.7,0.7}
\definecolor{forest}{rgb}{0.0,0.4,0.0}
\definecolor{grass}{rgb}{0.0,0.55,0.25}
\definecolor{olive}{rgb}{0.4,0.4,0.0}
\hypersetup{colorlinks=true, linkcolor=grass, citecolor=blue}

\begin{document}
\title{{\color{navy}Molecular-Orbital-Free Algorithm for Excited States\\ in Time-Dependent Perturbation Theory}}
\author{Melissa J. Lucero}
\email[]{lucero@lanl.gov}
\homepage[]{http://www.t12.lanl.gov/lucero}
\affiliation{Theoretical Division, Los Alamos National Laboratory, Los Alamos, New Mexico 87545}
\author{Anders M. N. Niklasson}
%\email[]{amn@lanl.gov}
%\homepage{http://www.lanl.gov/?}
% 
 \author{Sergei Tretiak}
\affiliation{Theoretical Division, Los Alamos National Laboratory, Los Alamos, New Mexico 87545}
\author{Matt Challacombe}
%\email[]{mchalla@lanl.gov}
%\homepage{http://www.lanl.gov/mchalla}
\affiliation{Theoretical Division, Los Alamos National Laboratory, Los Alamos, New Mexico 87545}
\begin{abstract}
A non-linear conjugate gradient optimization scheme is used to obtain excitation energies within the Random Phase Approximation (RPA).
The solutions to the RPA eigenvalue equation are located through a variational characterization using a modified Thouless functional, which is based upon an asymmetric Rayleigh quotient, in an orthogonalized atomic orbital representation.
In this way, the computational bottleneck of calculating molecular orbitals is avoided.
The variational space is reduced to the physically-relevant transitions by projections.
The feasibility of an RPA implementation scaling linearly with system size, $N$, is investigated by monitoring  convergence behavior with respect to the quality of initial guess and sensitivity to noise under thresholding, both for well- and ill-conditioned problems.
The molecular-orbital-free algorithm is found to be robust and computationally efficient providing a first step toward a large-scale, reduced complexity calculation of time-dependent optical properties and linear response. 
The algorithm is extensible to other forms of time-dependent perturbation theory including, but not limited to, time-dependent Density Functional theory.

\end{abstract}
\keywords{RPA; excited states; linear scaling; nonlinear conjugate gradient; variational; Thouless functional; asymmetric Rayleigh quotient; electronic structure; time-dependent perturbation theory; TDHF; TDDFT; reduced complexity; Liouville matrix}
\maketitle
\section{Introduction} %Section =======================================================
Matter responds to electromagnetic perturbation in a \emph{time-dependent} fashion: incident light induces periodic fluctuations within the electron density of a molecule that can be described by its excitation spectrum.
The excitation spectrum is of fundamental importance to many fields, ranging from analysis of interstellar clouds to the molecular basis of disease.
Unfortunately, excited states are difficult to calculate for large complex systems because the
scaling of the computational cost with the number of atoms is prohibitive. 
While numerous efforts have been devoted to the development of reduced complexity algorithms for ground state properties,\cite{Goedecker99} much less work has been focused on efficient algorithms for excited state response properties.
The purpose of this paper is to investigate a method for variational characterization of the excitation spectrum that could potentially scale linearly with system size.
This would allow studies of much larger systems than currently achievable.
The excitation spectrum is described by the Random Phase Approximation (RPA) within time-dependent Hartree-Fock theory,\cite{Dirac30,Frenkel34,Heinrichs68,Thouless72,Ring80,McWeeny89,Szabo96} but our algorithm is general and can be applied also to time-dependent Density Functional Theory.\cite{Runge84,Dreuw2005}
\subsection{The RPA Equation} %SUBSECTION -------------------------------------------------------------------------
Concomitant to the development of Many Body theory to describe the ground states of molecules, work to calculate properties of the more elusive excited states employing the Random Phase Approximation (RPA) began in the early 1950's. 
Avoiding the complications of addressing independent particles in a many electron system, the original, classical mechanical RPA treats the electron-repulsion terms as part of an ensemble average.
The Fourier transforms of the Coulomb terms have ``random phases'' that cancel, hence the name.\cite{Bohm51, Pines52} 
Recognition of an explicit quantum mechanical connection to single determinants\cite{Nozieres58} led to the demonstration of equivalence between the RPA and a time-dependent extension of Hartree-Fock (HF) theory --~thus permitting a fully quantum mechanical treatment of matter under light-induced perturbation.\cite{Ehrenreich59} 

During the 1960's, three equivalent formalisms developed around application of the RPA to calculate excited states.
While these derivations, based upon Equations of Motion,\cite{Zwanzig64, Sawada57b, Baranger60}
Green's functions,\cite{Ferrell57b, Thouless61} and time-dependent Hartree Fock theory\cite{Ferrell57a, Goldstone59, Rowe66b} are non-trivial, \cite{Ring80,McWeeny89,Cook2005} it is sufficient to note that for electronic transitions, a description utilizing only the one-body density is valid \textendash ~provided that the particle excitation energies are smaller than the Fermi energy and two-body correlations can be neglected.\cite{Ferrell57a} 
In this case, electronic excitations are well-described by the RPA eigenvalue equation,\cite{Cook2005,Ring80} written in the Molecular Orbital, ``MO'' basis as
  \begin{equation}\label{eq:RPA} %EQN RPA
  \begin{pmatrix} \bm A & \bm B \\ \bm B^* &\bm A^*\end{pmatrix}\binom{\vec X}{\vec Y} = \omega \begin{pmatrix} \bm I & 0 \\ 0 & -\bm I\end  {pmatrix} 
  \binom{\vec X}{\vec Y} ~.
  \end{equation}
The resonant frequencies, or excitation energies, are represented by the eigenvalues, $\omega$.
The elements of the matrices \mtrx A and \mtrx B are given by
\begin{equation}\label{eq:elementsA} %EQN Square matrices A and B
  A_{mi,nj} = (\epsilon_{m} - \epsilon_{i}) \delta_{ij}\delta_{mn} + V_{mj,in} -V_{mj,ni} \,,
\end{equation}
and
 \begin{equation}\label{eq:elementsB}
  B_{mi,nj} =   V_{mn,ij} -V_{mn,ji} ~,
\end{equation}
and the elements of $V_{mn,ij}$ are the conventional two-electron integrals.\cite{Szabo96}
The $i,j$ indices are from the set of occupied states while the $m,n$ indices correspond to the virtual orbitals, while $\epsilon_{m}$ and $\epsilon_{i}$ denote the Fockian eigenenergies.

The matrices \mtrx A and \mtrx B correspond to 4th order tensors of dimension ($N_{occ}\times N_{virt}) \times (N_{virt}\times N_{occ}$) spanning the Liouville space of transitions between the occupied (\emph{occ}) and virtual (\emph{virt}) subspaces.
These act upon the vectors $\vec X$ and $\vec Y$, composed of orbital coefficients, so that particle-hole (\emph{ph}) transitions are described by $\vec X$ while $\vec Y$ contains the hole-particle (\emph{hp}) transitions. 

The first term of \mtrx A in Eq.~(\ref{eq:elementsA}) corresponds to the undressed, bare excitations, i.e., those predicted by Koopman's theorem. 
The last two terms, or \mtrx B, add a correlation-based correction to the bare energies of \mtrx A  based upon the Coulomb and exchange interactions.
(Setting  $Y=0$, produces the Tamm-Dancoff\cite{Tamm91,Tamm45, Dancoff50} approximation.)
Finally, the unitary matrix, 
$ \mathds{N}=( \begin{smallmatrix}
       \bm I & \bm 0\\
       \bm 0 & -\bm I \end{smallmatrix} \bigr)$,
is a unit diagonal metric tensor, serving as an orthonormalization constraint,\cite{Thouless61,Thouless61a} defining the indefinite inner product associated with the space of \emph{ph-hp} transitions, in the Molecular Orbital basis.
\subsection{Linear Scaling Approaches to Solving the RPA Equation} %SUBSECTION ------------------------
The RPA equation was originally derived in the molecular orbital representation, as in Eq.~(\ref{eq:RPA}), and the familiarity of ``molecular orbitals'' in discussions involving ground states render it a popular basis in which to work.\cite{Appel2003,Dreuw2005}
However, the molecular-orbital representation requires a full eigenfunction solution of the ground state problem, which typically requires a computational cost that scales as ${\cal O}(N^3)$, where $N$ is the number of basis-functions, assumed to be proportional to system size.
A requirement for any reduction of this computational ${\cal O}(N^3)$ bottleneck is, therefore, to find a \emph{molecular-orbital-free} algorithm for the solution of the RPA equation. 

Recently, a number of groups \cite{Coriani2007,Izmaylov2006,Xiang2006,Weber2005,Yam2003,Larsen2000} have achieved a linear scaling computational complexity for the ground state self-consistent field (SCF) problem in Hartree-Fock (or density functional theory) using ``fast'' algorithms for computation of the Fockian $\bm F$ and sparse matrix algebra (dropping of small elements) to exploit quantum locality of the density matrix $\bm P$. 
If the transition densities in the time-dependent response equations also demonstrates quantum locality, then the same fast methods used for the ground state problem are applicable.\cite{Yam2003,Izmaylov2006} 

Solving the time-dependent quantum response problem in ${\cal O}(N)$ is pivotal in studies of large scale systems currently inaccessible to conventional methods.
Perhaps the most successful approach, to date, is to propagate an electron impulse response through numerical integration in real time, \cite{Nomura97,Yokojima99,Iitaka2000,Yokojima2000} and then retrieve the spectra from the time series through Fourier transformation. 
More recently, Coriani, {\em et al.}\cite{Coriani2007} have implemented the matrix exponential approach of Larsen, {\em et al.}, \cite{Larsen2000} observing an acceleration in computation of excitation energies for one dimensional systems. 

Another reduced complexity approach for time-dependent response calculations was recently presented by Izmaylov, {\em et al.} \cite{Izmaylov2006} and Kussman and Ochsenfield.\cite{Kussmann2007}
It is worth noting that in the adiabatic zero-frequency limit, when $\omega\rightarrow0$, the adiabatic response problem can be solved with surprising efficiency in linear scaling complexity using adiabatic density matrix perturbation theory based on purification.\cite{Niklasson2004}
Linear scaling density matrix perturbation theory can be applied to the calculation of response properties of molecules, both for lower \cite{Weber2004} and higher order perturbations, \cite{Weber2005} as well as for the crystalline problem, \cite{Xiang2006} including the electric polarizability.

The reduced complexity approach in this paper is based upon a well-established variational characterization of the eigenvalue spectrum as applied to the RPA equation.
The key idea is to use an molecular-orbital-free approach, avoiding the ${\cal O}(N^3)$, bottleneck. This is achievable through a functional optimization of an asymmetric Rayleigh Quotient as formulated by Thouless more than four decades ago.\cite{Thouless61}
The intent of this paper is not to present a linear scaling algorithm, but to analyze and discuss 
the limitations and feasibility of a variational optimization of a Thouless functional in the context of reduced complexity calculations.
\section{Molecular-orbital-free time-dependent perturbation theory}\label{nomo} %SECTION =========
To derive a molecular-orbital-free formulation for the RPA equation suitable to  ${\cal O}(N)$  calculations, we may start from time-dependent Hartree-Fock theory,\cite{Cook2005,Ring80}
\begin{equation}
i \frac{\partial \bm{P}}{\partial t} = [\bm F,\bm P]_{\scriptscriptstyle {\bm S}} = \bm{FPS} -\bm{SPF},
\end{equation}
where $\bm S$ is the overlap matrix, $\bm P$ is the single-particle density matrix of the Hartree-Fock ground state, and $\bm F$ is the effective single-particle Hamiltonian, i.e., the Fockian (or the Kohn-Sham Hamilitonian, in a generalization to time-dependent Density Functional theory). In an orthogonalized representation, $\bm S$ becomes the identity matrix.

Looking at the first-order response under variation of the density matrix $\delta \bm P$, we find that, 
\begin{equation}
i \frac{\partial \delta \bm{P}}{\partial t} = [\bm F, \delta \bm P]_{\scriptscriptstyle {\bm S}} + \left [\bm G(\delta \bm P),\bm P \right ]_{\scriptscriptstyle {\bm S}},
\end{equation}
which, in the frequency domain, gives the RPA linear response eigenvalue equation,
\begin{equation}\label{eq:lre}
 [\bm F,  \bm x]_{\scriptscriptstyle {\bm S}} + \left [\bm G( \bm x),\bm P \right ]_{\scriptscriptstyle {\bm S}}=\omega \bm x.
\end{equation}
Here, $\bm x$ is the Fourier transform of $\delta \bm P$ and
\begin{equation}\label{eq:alpha}
\bm G(\bm x) = 2 \bm J[\bm x] - \bm K [\bm x].
\end{equation}
The left commutator in Eq.~(\ref{eq:lre}) {gives the zeroth-order approximation corresponding the bare excitations, and the second commutator with $\bm G(\bm x)$,  includes additional Coulomb, $\bm J[\bm x]$, and exchange screening, $\bm K [\bm x]$.
In a generalization to time-dependent DFT, the exchange screening is replaced by the exchange correlation screening, i.e., the second-order functional derivative of the exchange-correlation action. \cite{Hohenberg64,Kohn65,Runge84,Petersilka96} 
The RPA excitation spectrum is thus given by the eigenfrequencies $\omega$ corresponding to \emph{ph,hp} transitions in Eq.~(\ref{eq:lre}). 

In a compact form, we can express the RPA equation, Eq.~(\ref{eq:lre}) as 
\begin{equation}\label{eq:compact} %EQN Compact RPA
\mathds{L} {\vec x} = \omega \vec x.
\end{equation}
\noindent The vector ${\vec x}$ is dyadic, corresponding to the unrolled $N \times N$ matrix $\bm x$, i.e.,
${\bm x}_{\scriptscriptstyle N\times N} \Leftrightarrow {\vec x}_{\scriptscriptstyle
N^2\times 1}$, where the double-headed arrow  denotes both equivalence and a tensorial mapping, or simply a stack operation.\cite{Graham81} 
For the matrix transpose $\bm x^{T}$, we use the corresponding unrolled vector notation $\vec x^{t}$.
In the following, we employ a mixed super-vector/matrix notation; projection is most natural for matrices, while the use of a vector notation lends itself to gradient-based minimization.
The action of $\mathds{L}$ onto $\vec x$ in Eq.(\ref{eq:compact}) is thus given by 
\begin{equation} \label{eq:general}
\mathds{L} \vec x  \Leftrightarrow [\bm F, \bm x]_{\bm \scriptscriptstyle{S}} + [\bm G(\bm x), \bm P]_{\bm \scriptscriptstyle{\bm S}}.
\end{equation}

The general formulation of the RPA equation,  as expressed in Eqs. (\ref{eq:lre}) and (\ref{eq:compact}), is independent of the basis-set representation and can thus be applied in a molecular-orbital-free approach, avoiding an expensive diagonalization scaling as $\mathcal O (N^{3})$ with system size $N$.
In our molecular-orbital-free algorithm, we employ an orthogonalized atomic orbital basis representation, with $\bm S=\bm I$, which can be efficiently constructed with $\mathcal O (N)$ complexity through a congruence transformation,\cite{Golub83} e.g., based upon the approximate inverse Cholesky transform.\cite{Benzi2001}
While this work involves solution of the RPA eigenvalue equation, extension to time-dependent Density Functional theory is straightforward.
\section{Non-Linear Conjugate Gradient Optimization of the Thouless Functional}%=================
While the Lanczos algorithm has been used to iteratively solve the RPA equation,\cite{Tretiak2002} severe problems are experienced when calculating high-lying excitations, e.g., due to orthogonality constraints.\cite{Bertsch2002,Fabrocini2002} 
More importantly, achieving linear scaling complexity requires sparse linear algebra, which may preclude the Lanczos algorithm due to numerical instabilities.\cite{vandenEshof2004b, Simoncini2002,vandenEshof2005,Simoncini2007} 
Our molecular-orbital-free scheme utilizes a non-linear conjugate gradient optimization of a Rayleigh quotient related to the method of Muta.\cite{Muta2002} 
The use of non-linear conjugate gradients are particularly advantageous in the context of linear scaling algorithms, because of the ability to remain robust under an incomplete sparse matrix algebra, as demonstrated in the work of Simoncini\cite{Simoncini2002} and Notay.\cite{Notay2003}

The core of our algorithm is a variational characterization of the excitation spectrum based on the Thouless functional.\cite{Thouless61}
Thouless demonstrated the possibility of a variational approach to solving the RPA equation \emph{via} iterative optimization of an asymmetric Rayleigh Quotient.
This functional, when expressed in representation-independent form, becomes 
\begin{equation}\label{eq:Thouless2} %EQN Thouless GENERAL
\Omega[\vec x]  = \frac{ {\vec x}^{\scriptscriptstyle t} \cdot \mathds{L}  {\vec x}_{\scriptscriptstyle{\mathds{N}}}}{\abs{{\vec x} \cdot {\vec x}_{\scriptscriptstyle{\mathds {N}}}}},
\end{equation}
where it is understood that the numerator is computed as in Eq.~(\ref{eq:general}) and the denominator is given by the absolute value of the Euclidean vector product denoted by the dot between $\vec x$ and $\vec x_{\scriptscriptstyle{\mathds{N}}}$.
The metric tensor $\mathds{N}$ is included through
\begin{equation}
\vec x_{\scriptscriptstyle{\mathds{N}}} =  \mathds{N} \vec x \Leftrightarrow (\bm P - \bm Q) \bm x~,
\end{equation}
where the subscript $\mathds{N}$ denotes the action of $\mathds{N}$ onto $\vec x$, $\bm P$ is the occupied subspace projector and  $\bm Q=\bm I -\bm P$ is the complimentary projector for the virtual subspace.

Only stationary solutions to the Thouless functional in Eq.~(\ref{eq:Thouless2}), corresponding to \emph{ph-hp} transitions between the occupied and virtual subspaces, are of physical relevance. 
Rather than impose  \emph{ph-hp} symmetry explicitly by construction, it is straightforward to reduce the variational search space to the physically-relevant solutions by the projection.
\cite{Zwanzig60,Karplus63,Langhoff72,McWeeny89}
\begin{equation}\label{eq:PvQ}%EQN PvQ
 \bm{x}_{\scriptscriptstyle{\bm P}}= \bm P\bm x \bm Q + \bm Q \bm x  \bm{P}
\end{equation}
or, equivalently,
\begin{equation}\label{eq:PvQ2}%EQN PvQ2
 \bm{x}_{\scriptscriptstyle{\bm P}}= [[\bm x,\bm P], \bm P]~.
 \end{equation}
This projection conforms to the \emph{ph-hp} formalism of the RPA Equation in the MO basis, Eq.~(\ref{eq:RPA}), with removal of non-physical states and reduces the size of the variational search space considerably.

For large, sparse problems, it is possible to construct the $\bm P$ and $\bm Q$ projectors (or density matrices) with linear scaling complexity using recursive purification methods.\cite{McWeeny56,Palser98,Niklasson2002}
In the general case, the metric tensor $\mathds{N}$, which occurs implicitly in the Thouless functional, corresponds to the indefinite scalar product \cite{Thouless62,Ring80a}
\begin{equation}\label{famnorm} %EQN Norm
(\vec v,\vec u)_{\scriptscriptstyle{\bm P}}   \Leftrightarrow  \mathrm{Tr}\lbrace \bm v^{\scriptscriptstyle{T}}_{\scriptscriptstyle{P}} \, [\bm u_{\scriptscriptstyle{p}}, \bm P] \rbrace = \mathrm{Tr}\lbrace \bm v^{\scriptscriptstyle{T}}_{\scriptscriptstyle{P}} \bm (\bm P - \bm Q)\bm u_{\scriptscriptstyle{p}} \rbrace
\end{equation}
with the norm 
\begin{equation}\label{eq:norm2}
\norm{\bm x}_{\scriptscriptstyle {\bm P}} = \norm{\vec x}_{\scriptscriptstyle {\bm P}}  = \sqrt{\abs{(\vec x,\vec x)_{\bm \scriptscriptstyle{P}} } }~.
\end{equation}
We may now consider optimization of the representation-independent Thouless functional,
\begin{equation}\label{eq:anders}
\Omega[\vec x]=\frac{(\vec x, \mathds L \vec x)\scriptscriptstyle{\bm P}}{\norm{\bm x}_{\scriptscriptstyle {\bm P}}^{\scriptscriptstyle{2}}}~.
\end{equation} 
\noindent This formulation of the Thouless functional implicitly invokes \emph {ph-hp} symmetry for the excitations, which produces paired eigenvalues, $\pm \omega_{i}$ as would be associated with the Liouville operator.\cite{Antoniou98}

To locate the first  transition, $(i=1)$, we minimize  Eq.~(\ref{eq:anders}),
\begin{equation}
\vec v_i = \argmin_{\vec x} \Omega[\vec x]
\end{equation}
which yields the eigenfrequency 
\begin{equation}
\pm \omega_i= \pm \Omega[\vec v_i] ~.
\end{equation}
The search for  subsequent eigenstates requires that lower lying eigenvectors be either projected out or shifted away, such that they are not rediscovered by consecutive minimization.
We use a Wilkinson shift\cite{Wilkinson65} of the interior eigenvalues, shifting $\omega_{j}$ to  $\omega_{j} + \sigma$,  which is 
outside the region of interest, written as the shifted $\mathds{L}\vec x$:
\begin{equation}\label{eq:FoldShift} %EQN Fold Shift
\mathbb{L} \,\vec{x} + \sum_{j}^{i} \left(\omega_{j}+ \sigma \right) \left[ \vec{v_{j}}(\vec{v}_{j}, \vec{x})_\mathbf{\scriptscriptstyle {\bm P} }  + \vec{v}^{\scriptscriptstyle \, t}_{j}(\vec{v}^{\scriptscriptstyle \, t}_{j}, \vec{x})_\mathbf{\scriptscriptstyle {\bm P} } \right],
\end{equation}
where $\{ \vec v_{j} \}$ and $\{ \vec w_{j} \}$ are  previously determined RPA eigenstates and excitation energies, respectively.

Our molecular-orbital-free algorithm utilizes a conventional Polak-Ribi\`{e}re nonlinear conjugate gradient algorithm,\cite{Polak69,Polyak69} with restarts, in an orthogonal atomic orbital (e.g., L\"owdin\cite{Lowdin50}) basis, as is summarized in Fig.~\ref{code}. The outer loop runs over the first $M$ interior eigenvalues, while the inner loop iterates over the non-linear conjugate gradient steps. 
The projections in Lines 4 and 10 eliminate non-physical states by imposing \emph{ph, hp} symmetry, which significantly reduces the search space. 
Note: the action of the  $\mathds L$ operator is performed through the Fock builds in Eq.~({\ref{eq:general}).
The first $\mathds L$ operation on Line 7 is used for calculation of the gradient, and the second  $\mathds L$ operation occurs at Line 13, which is used in the line search of Line 14.

The Wilkinson shift occurs on Lines 6 and 7 while construction of the Thouless functional occurs in Line 8. 
The gradient $\vec g_{\scriptscriptstyle}$, is defined in Lines 9 and 10 and the the conjugate gradient search directions $\vec p_{\scriptscriptstyle k}$ are given by the subsequent calculations on Lines 11 and 12.
Restarts for $\beta = 0$ were not necessary, and did not occur during our test calculations.

The functional minimum of the line search along the conjugate gradient directions on Line 14 is given by 
\begin{equation}\label{eq:quadfn}
\lambda_\pm  = \frac{-b \pm \sqrt{b^2-4ac}}{2ac} 
\end{equation}
where
\begin{multline}
 a = ( \vec p , \vec t )_{\scriptscriptstyle {\bm P}} \left [( \vec p , \vec x )_{\scriptscriptstyle {\bm P}} +
 ( \vec x , \vec p )_{\scriptscriptstyle {\bm P}} \right]\\
-( \vec p , \vec p )_{\scriptscriptstyle {\bm P}} \left[( \vec p , \vec s)_{\scriptscriptstyle {\bm P}} +
 ( \vec x , \vec t )_{\scriptscriptstyle {\bm P}} \right]  
  \end{multline}
 \begin{equation} 
 b  =   2 ( \vec p , \vec t )_{\scriptscriptstyle {\bm P}} ( \vec x , \vec x )_{\scriptscriptstyle {\bm P}}
 -2( \vec x , \vec s)_{\scriptscriptstyle {\bm P}} ( \vec p ,  \vec p )_{\scriptscriptstyle {\bm P}} \qquad \qquad \quad ~
\end{equation}
\begin{multline}
c = ( \vec x ,  \vec x )_{\scriptscriptstyle {\bm P}} \left[( \vec p , \vec s)_{\scriptscriptstyle {\bm P}} +
( \vec x , \vec t )_{\scriptscriptstyle {\bm P}} \right] \\
-( \vec x , \vec s)_{\scriptscriptstyle {\bm P}} \left[( \vec p ,  \vec x )_{\scriptscriptstyle {\bm P}} +
 ( \vec x , \vec p )_{\scriptscriptstyle {\bm P}} \right] .
\end{multline}
After each inner loop iteration over $k$, the desired $i$th eigenpair composed of eigenvector the $\vec v_{\scriptscriptstyle i}$ and eigenvalue $\omega_{\scriptscriptstyle i}$, are given on Lines 17 and 18.
\begin{figure} % FIGURE: PSEUDOCODE (Figure 1.)
\caption{The Molecular-Orbital-Free RPA Algorithm.\\ \label{code}}
\NumberProgramstrue
\sfvariables
\begin{programbox}
 \FOR i=1 \TO M \DO
  Generate ~initial ~vector, \displaystyle \vec{x}
 \FOR k = 1, until ~convergence
   \displaystyle \bm{x}= \mathbf{P} \bm{x}  \mathbf {Q} + \mathbf{Q }\bm{x}  \mathbf{P}^{\phantom{\displaystyle \sum}}_{\phantom{\displaystyle \sum}}
   \displaystyle \vec{x}= \frac{\vec{x}} {\norm{\vec{x}}_\mathbf{\scriptscriptstyle P} }_{\phantom{\displaystyle \sum}}
   \displaystyle \vec {s}=\sum_{j}^{i-1} \left(\omega_j+\sigma)  \right) \left[ \vec{v_{j}}(\vec{v}_{j}, \vec{x})_\mathbf{\scriptscriptstyle {\bm P} }  + \vec{v}^{\scriptscriptstyle \, t}_{j}(\vec{v}^{\scriptscriptstyle \, t}_{j}, \vec{x})_\mathbf{\scriptscriptstyle {\bm P} } \right]
%s STEP 7   
      \displaystyle \vec{s}=\Liou \vec{x} + \vec s_{\phantom{\displaystyle \sum}}^{\phantom{\displaystyle \sum}}
      \displaystyle \Omega = \frac{(\vec{x}, \vec{s})_{\scriptscriptstyle  \bm P}} {\norm{\vec x}^{2}_{\scriptscriptstyle {\bm P}}}_{\phantom {\displaystyle \sum} }
%   \displaystyle \epsilon = \left\arrowvert {\mid \Omega \mid -\max_{j=1\dots N^{2}} \left( \frac{ \mid \vec{s}_{\scriptscriptstyle j} \mid }{\mid \vec{v}_j \mid } \right)}\right\arrowvert ~  \IF \epsilon < \eta ~ \EXIT , \FI_{\phantom {\displaystyle \sum} }
   \displaystyle \vec{g}_k = 2\vec s - 2\Omega \vec x^{\phantom {\displaystyle \sum} }_{\phantom {\displaystyle \sum} }
% STEP 10
   \displaystyle \bm{g}_k= \mathbf{P}  \bm{g}_k \mathbf {Q} + \mathbf{Q }  \bm {g}_k \mathbf{P}_{\phantom {\displaystyle \sum} }
% -PR- Beta
\displaystyle \beta = max\left\{\frac{(\vec{g}_{\scriptscriptstyle k} - \vec{g}_{\scriptscriptstyle{k-1}}, \vec{g}_{\scriptscriptstyle{k}}) _{\scriptscriptstyle{\bm P}}} {\norm{\vec g_{\scriptscriptstyle {\bm k-1}}}^{2}_{\scriptscriptstyle{\bm P}}},0\right \}_{\phantom {\displaystyle \sum}}
   \displaystyle \vec p_{\scriptscriptstyle {k}} = \vec g_{\scriptscriptstyle {k}} + \beta \vec p_{k}   
   \displaystyle \vec t = \mathds L \vec p_{k} ^{\phantom {\displaystyle \sum} } 
      \displaystyle \lambda_k= \argmin_{\lambda} \Omega\lbrack \vec x + \lambda \vec p_{\scriptscriptstyle {k} } \rbrack^{\phantom {\displaystyle \sum} }
   \displaystyle \vec{x} = \vec{x} + \lambda_{\scriptscriptstyle{k}} {\vec p_{\scriptscriptstyle {k} }}^{\phantom {\sum}}
 \END\DO
   \displaystyle \vec{v}_{\scriptscriptstyle{i}}= \vec{x}^{\phantom { \sum}}
   \displaystyle  \omega_i =  \Omega [\vec{v}_i]
\END\DO
\end{programbox}
\end{figure}
\section{Performance of the Molecular-Orbital-Free Algorithm} %===============================
\subsection{Illustration of the RPA Eigenvalue Spectrum}% -------------------------------------------------------------
To illustrate the performance of the molecular-orbital-free solution of the RPA equation, the properties of the solutions and various relevant concepts, a  schematic picture containing a hypothetical set of spectra is provided in Fig.~\ref{fig:hypo}.

The spectrum to the extreme left, labeled (FULL), depicts a complete eigenvalue spectrum.
All eigenvalues, physical and non-physical are included: no projections have been performed.
There are numerous ``bands'' that might imply clustering or degeneracies which would slow down a variational search of the eigenstates.

The next spectrum, to the immediate right, (TRANSITIONS), demonstrates the effect of 
the projection in Eq.~(\ref{eq:PvQ}) including only the subspace of \emph{ph,hp} transitions.
The removal of unphysical states significantly reduces the density of eigenvalues, particularly around zero.
This strongly facilitates the search for eigensolutions of the RPA equation, in particular for low-lowing excitations of physical interest.\cite{Lehoucq2000,Verhaar60,Sleijpen2003,vandenEshof2004a,Li2004} 
\begin{figure}[!htb] %FIGURE  2
\includegraphics[width=3.4in]{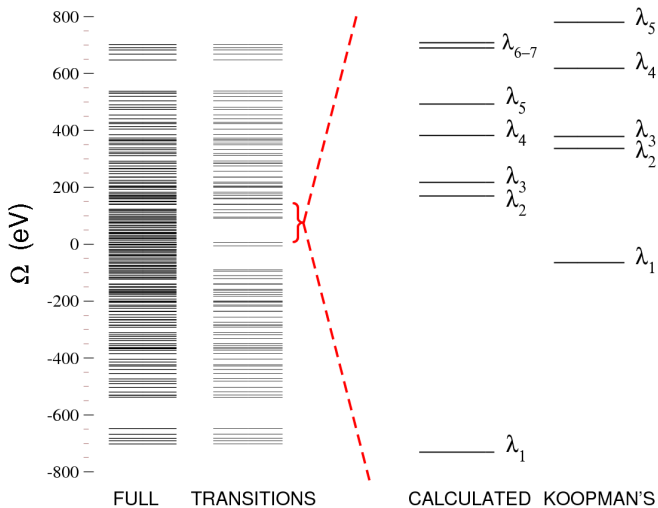}
\caption{\small\small Illustration of a set of eigenvalue spectra for the RPA equation. 
The entire spectrum, (FULL), is shown on the far-left. Notice the dense clustering of eigenvalues around zero.
Immediately to the right, (TRANSITIONS), depicts the spectrum containing 
only physical particle-hole and hole-particle excitations.  
The next spectrum, (CALCULATED), is an expansion emphasizing a few low-lying transitions.
The spectrum to the extreme right, (KOOPMAN's) contains the corresponding transitions as would be
estimated from Koopman's theorem. \label{fig:hypo}}
\end{figure}

The third and fourth spectra, on the right, CALCULATED and KOOPMANS, provide enlargements of the eigenvalue spectrum, in the positive region near zero.
The CALCULATED spectrum, shows low-lying \emph{ph-hp} transitions, given by the RPA eigenvalue equation.
Note that high energy excitations are not expected to be well-described by the RPA equation.\cite{Brown61,Brown64,Rowe68a,Cook2005} 
The Koopman's spectrum shows the bare excitations, without Coulomb or exchange screening, and corresponds to the eigenvalue excitation spectrum of the ground state Fockian.
Koopman's theorem provides only an approximate spectrum,\cite{Szabo96} but, as will be shown below, initial vectors derived from Koopman's theorem produce good initial guesses for our non-linear conjugate gradient optimization.
\subsection{Convergence Behavior }\label{examples}% SUBSECTION ---------------------------------------------
To study the behavior of our molecular orbital-free algorithm, it was prototyped in an orthogonalized atomic orbital basis employing dense linear algebra and a conventional $\cal O(N^4)$ approach to Hartree-Fock theory.
Our test calculations are thus not performed with any linear scaling complexity.
The present implementation is limited to $s$-type STO-3G basis functions and is not expected to produce chemically-relevant data.
Rather, we utilize this description to characterize the most important features related to more accurate representations, as well as consider the problems inherent to linear scaling implementation.
In this manner, we can simulate the influence of large basis sets, extended periodic systems and complex molecules.

As with more conventional approaches to solving the excitation problem within time-dependent perturbation theory, the work required to resolve an eigenstate increases with an increasing condition number,
$\kappa$, i.e., the ratio between the highest and lowest singular value of $\mathbb{L}$.  
To study convergence and other properties, we generated test systems using a linear arrangement of fourteen hydrogen atoms.
Progressively smaller inter-atomic spacings produce correspondingly higher condition numbers, $\kappa$, which parallels that which could arise as the size of system or basis set increases.
Notice, a variation of the condition number also illustrates the effect of preconditioning.
All condition numbers, $\kappa$, are estimated approximately in order of their magnitude.

To simulate the effects of an incomplete, sparse matrix algebra \textendash ~an absolute necessity for linear scaling capability \textendash ~we add a random matrix with elements of amplitude $\pm \tau$ after each application 
of $\mathds L$ to a vector, i.e., the $\mathbb{L} \vec{x}$ Fock builds.  
This is equivalent to using a looser numerical threshold in the case of a vanishing difference density matrix, which has been shown to yield linear scaling. \cite{Challacombe97} 

Two types of initial guesses are considered; a random guess, and a ``Koopmans' guess.''
The Koopmans' guess was based on direct diagonalization of the ground state Fockian.
This is an expensive procedure that certainly does not scale linearly with system size. 
The purpose is to study the effect of an improved initial guess.
An efficient ${\cal O}(N)$ construction of an accurate initial guess remains a very important, yet unsolved problem not discussed in this article.\cite{Goddard73,Huzinaga77,Flindt2004,Chassaing2005}

The convergence is measured in terms of the relative errors of the approximate RPA eigenvalues, $\epsilon_n$, where the error, 
\begin{equation} \label{eq:err}
\textrm{Err(n)}=\abs{\frac{{\epsilon_{\scriptscriptstyle{n}}-\omega_{\scriptscriptstyle{ref}}}}{\epsilon_{\scriptscriptstyle{n}}}}, 
\end{equation}
is calculated in each conjugate gradient iteration, $n$. 
The reference eigenvalues, $ \omega_{\scriptscriptstyle{ref}}$, were obtained from direct diagonalization of the $\mathds L$ matrices using the ZGEEV routine in the LAPACK library.\cite{Lapack2006} 

\subsubsection{Small $\kappa$, Varied Initial Guess}%~~~~~~~~~~~~~~~~~~~~~~~~~~~~~~~~~~~~~~
The convergence behavior for the first 5 eigenvalues, $\lambda_{1}-\lambda_{5}$, of the system with a condition number $\kappa = 10^{2}$ is depicted in Fig.~\ref{fig:fig3}.
The algorithm starts with randomly generated initial vectors for each of the 5 eigenvalues sought. The non-linear conjugate gradient optimization for each eigenvalue is then allowed to proceed until $\textrm{Err}(n) \lt 10^{-12}$.
\begin{figure}[!htb]  % FIGURE 3
\includegraphics[width=3.3in]{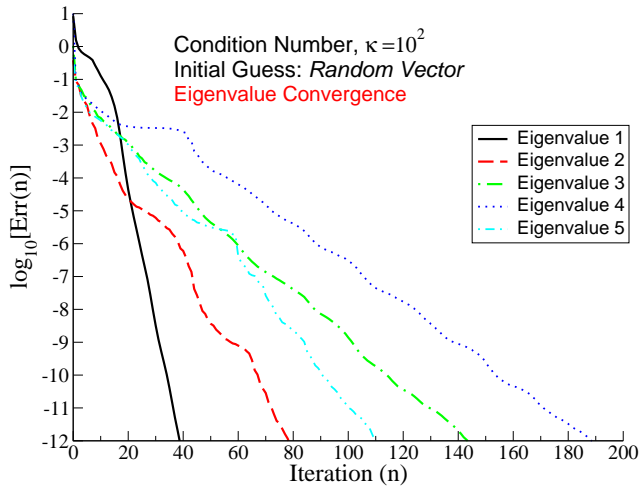}
\caption{\small Convergence of the first 5 eigenvalues for a well-conditioned ($\kappa=10^{2}$) 
matrix using random initial guesses. \label{fig:fig3}}
\end{figure}

Contrast these results with the curves in Fig.~\ref{fig:fig4}, which contains the convergence patterns for 
the same system, but this time starting with initial guess vectors based upon Koopman's theorem.  
Despite the apparent non-ideality implied by the hypothetical Koopman's-based spectrum (KOOPMAN's) in Fig.~\ref{fig:hypo}, the initial eigenvectors generated by Koopman's guess provide significant improvement in the rate of convergence compared to that of the random vectors used to generate the plots in Fig.~\ref{fig:fig3}.
For the third eigenvalue, the convergence is improved by almost an order of magnitude.
\begin{figure}[!htb]
\includegraphics[width=3.3in]{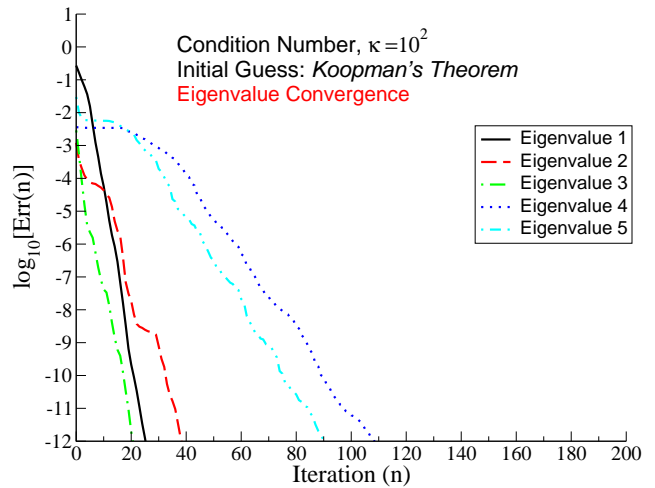}
\caption{Convergence of the first 5 eigenvalues for a well-conditioned ($\kappa = 10^{2}$) matrix using initial guesses derived from Koopman's theorem.
\label{fig:fig4}}
\end{figure}
\subsubsection{Convergence: Varied $\kappa$, Good Initial Guess} %~~~~~~~~~~~~~~~~~~~~~~~~~~~
While the behavior of the algorithm for well-conditioned matrices is useful for proof-of-concept, the performance of any algorithm in the presence of ill-conditioned matrices is of paramount importance for many problems, especially in the limit of large basis sets.
Figure \ref{fig:fig5} illustrates the convergence for condition numbers ranging from $\kappa=10^{1}$  to $10^{4}$. 
We use initial vectors based on Koopman's theorem in each case.
\begin{figure}[!ht]
\includegraphics[width=3.3in]{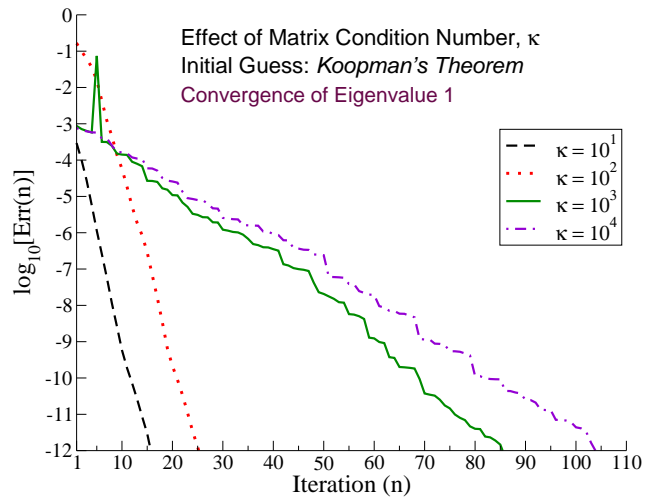}
\caption{ Convergence of the first eigenvalue for matrices with $\kappa \in [ 10^2, 10^4]$. In each case the Koopmans' guess was used.
\label{fig:fig5}}
\end{figure}
The curves in Fig.~\ref{fig:fig5} depict the convergence behavior for the lowest excitation energy 
for each condition number $\kappa$, but similar patterns are observed for all of the first 
five eigenvalues in each case. 
The dashed (black) and dotted (red) lines show convergence patterns for the more well-conditioned systems, whereas the solid (green) and dash-dotted (blue) lines represent convergence for the less well-conditioned systems. 
The small blip observed at iteration 8 for the $\kappa = 10^{3}$ system is an artifact of the error calculation because of a sign change relative to the reference value.

The number of iterations required to reach the convergence, $\textrm{Err}(n) \lt 10^{-12}$, increases by almost an order of magnitude when the condition number is increased.
This also indicates the potential improvement that could be reached by an efficient preconditioner.
For the two better-conditioned systems, the distribution of the smaller eigenvalues is more even, resulting in smooth curves and relatively rapid convergence. 
This pattern for the first eigenvalue is also observed in the more well-conditioned case in Figs.~\ref{fig:fig3} and~\ref{fig:fig4}.

In going from well- to ill-conditioned matrices, not only does the slope decrease, extending the number of iterations to convergence, but the morphology of the curves changes as well.
A pronounced step/plateau pattern is evident, particularly as the optimization proceeds.
This behavior is typical of conjugate gradient schemes with clustered eigenvalues. \cite{Luenberger84,Golub83,Nocedal2006}
\subsection{Sensitivity to Numerical Error} % SUBSECTION: -----------------------------------------------------------
To probe the robustness of the molecular-orbital-free algorithm, we added noise of varying 
levels to a well-conditioned ($\kappa = 10^{2}$) system, as illustrated in Fig. \ref{fig:fig6}. 
Again, we used a Koopman's theorem-based initial vector and observed the convergence behavior for the first eigenvalue.
\begin{figure}[!ht]
\includegraphics[width=3.3in]{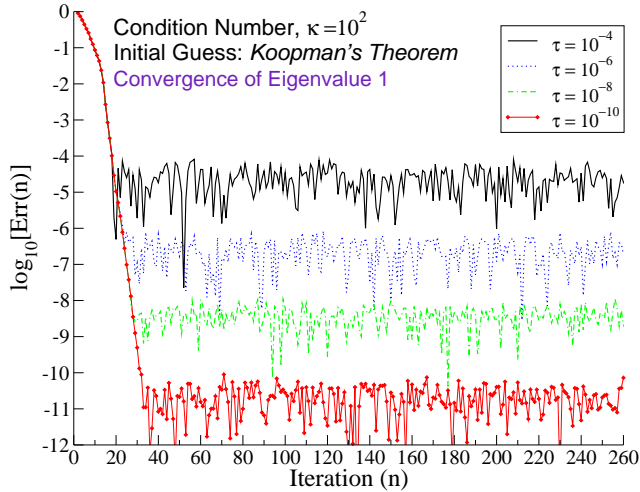}
\caption{ Convergence of the first eigenvalue for $\kappa  \sim 10^{2}$ with random noise in 
the range $\tau \in {\scriptstyle{\pm}} [ 10^{-8},10^{-4}]$, using initial guesses based upon 
Koopman's theorem.\label{fig:fig6}}
\end{figure}
Random noise in the range, $\tau \in {\scriptstyle{\pm}} [ 10^{-8},10^{-4}]$, provides a reasonable 
estimate of the induced errors we encounter in a typical linear scaling implementation, where small
elements below some chosen numerical tolerance are set to zero. 
We added this noise to every component of the newly formed vector ($\mathds{L} \vec{x}$) for each iteration of 
the inner conjugate gradient  loop (the ``'$k$'' loop) in order to simulate the accumulation of numerical 
error as the calculation proceeds (see Fig.~\ref{code}). 

We find that the algorithm is robust and stable with respect to numerical noise and that the error at convergence scales approximately linearly with the level of noise.
The same behavior is also observed for more ill-conditioned systems, as shown in Fig.~\ref{fig:fig7}.
\begin{figure}[!ht]
\includegraphics[width=3.3in]{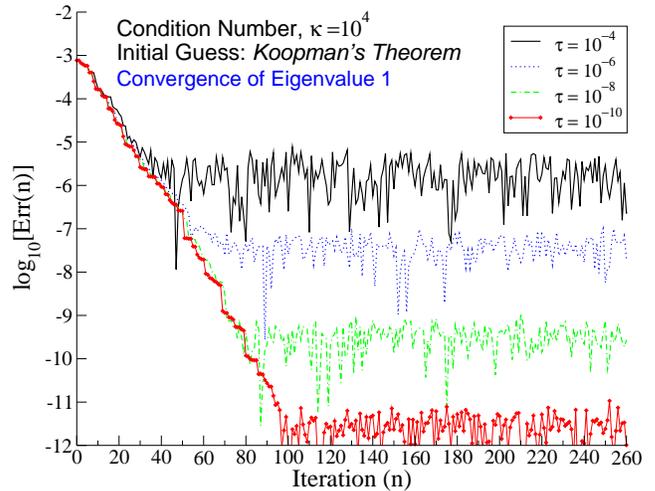}
\caption{ Convergence of the first eigenvalue for $\kappa \sim 10^{4}$ for  random noise in the range, 
$\tau \in {\scriptstyle{\pm}} [ 10^{-8},10^{-4}]$ using initial guess vectors derived from Koopman's theorem.\label{fig:fig7}}
\end{figure}
\section{Discussion and Conclusions} % SECTION =========================================
We have presented an algorithm for the variational characterization of the RPA eigenvalue spectrum, based on the Thouless functional and a non-linear conjugate gradient optimization in a molecular-orbital-free representation.
We have analyzed the convergence with respect to initial guess, condition number (preconditioning) and numerical noise.
This analysis clearly indicates a potential for reduced complexity calculations of large systems.
However, there remain several open questions: 1) The search space for the excitation spectrum corresponding to the dimensions of the Liouville operator $\mathds L$ in the RPA eigenvalue problem scales quadratically, $\mathcal O(N^2)$, with system size. 
Unless the initial guess is highly accurate, we can expect the number of iterations required to reach convergence to increase with system size. 
This would obviate linear scaling complexity. 
2) Unfortunately, the construction of a highly accurate initial guess is computationally very expensive. 
For example, building the Koopman's guess would typically require a diagonalization of the Fockian which scales as $\mathcal O(N^3)$.
A reduced complexity technique for finding a good initial guess or accurate preconditioning\cite{Bergamaschi2000,Simoncini2002,Notay2003,Simoncini2005,Simoncini2007} remains an unsolved, important challenge, though many efficient constructions may be possible. 
3) The stability of the Wilkinson shift under sparse linear matrix algebra has not been fully investigated, though the stability under noisy conditions indicates that this is not a problem.

In conclusion, the molecular-orbital-free scheme based upon a well-established variational characterization of the RPA excitation spectrum exhibits most of the necessary features required for an efficient linear scaling implementation. 
While further work remains, we believe this technique will become highly valuable for determination of large-scale excited state properties.
\section*{ACKNOWLEDGMENTS} %====================================================
We would like to express our gratitude to Professor C.~J.~ Tymczak and Drs. Richard L. Martin, Antonio Redondo, Kimberly W. Thomas, Eddy M. Timmermans and Val\'{e}ry Weber for many helpful discussions. M.J.L. gratefully acknowledges the support of a LANL Director's Postdoctoral Fellowship. This work was performed under the auspices of the U.S. Department of Energy and the Los Alamos National Laboratory LDRD program.

\bibliography{NLCG_RPA.bib}

\end{document}